\documentclass[%
 reprint,
 amsmath,amssymb,
 aps,
prb,
]{revtex4-2}
\usepackage{todonotes}
\usepackage{graphicx}
\usepackage{dcolumn}
\usepackage{bm}
\usepackage{caption}
\usepackage{subcaption}
\usepackage{inputenc}

\newcommand{\ignore}[1]{}

\newcommand{\zstepn}{h}
\newcommand{\zstep}{h}

\begin{document}

\preprint{APS/123-QED}

\title{Accurate computation of chemical contrast in field ion microscopy}

\author{Shalini Bhatt}
\author{Shyam  Katnagallu}
\author{J\"org Neugebauer}

\author{Christoph Freysoldt}

\affiliation{Max-Planck-Institut  f\"ur  Eisenforschung GmbH, D\"usseldorf, Germany
}%

\begin{abstract}
We present a computational approach to simulate
local contrast observed in Field Ion Microscopy (FIM). It is based on density-functional theory utilizing the Tersoff-Hamann approach as done in Scanning Tunneling Microscopy (STM). A key requirement is the highly accurate computation of the surface states' wave-function tails.
To refine the Kohn-Sham states from standard iterative global solvers we introduce and discuss the {\bf EX}trapolation of \textbf Tails via \textbf Reverse integration \textbf Algorithm (EXTRA). The decaying tails are obtained by reverse integration (from outside in) using a Numerov-like algorithm. The starting conditions are then iteratively adapted to match the values of plane-wave Kohn-Sham wave functions close to the surface. We demonstrate the performance of the proposed algorithm by analysing and showing the chemical contrast for Ta, W, and Re at Ni surface.

\end{abstract}

\maketitle
\section{\label{sec:level1}Introduction}
Field ion microscopy (FIM) was the first microscopy technique to image individual atoms on a metal surface with atomic spatial resolution\cite{muller1951feldionenmikroskop,muller1956resolution}. An imaging gas (e.g. He, Ne) is ionized above the surface of a nano-sharp needle-shaped specimen (end diameter $<100$ nm) subject to a high electric field of $10^{10}$ V/m.
The ions are then accelerated along the field lines on to a
two-dimensional detector to produce an image of the tip surface magnified \cite{brandon1963resolution} by a factor of 10$^7$.
While FIM has been mostly replaced by scanning
tunneling microscopy \cite{chen2021introduction}, atomic force microscopy, and similar scanning probe techniques\cite{lucier2005determination} for surface characterization, there has been a recent resurgence when combined
with field evaporation. Field evaporation refers to the process by which electric fields induce the removal of atoms from needle-shaped specimens. This technique enables FIM to extend its capabilities to provide atomically resolved 3D characterization. The ions produced by field evaporation can be identified using time-of-flight spectrometry, which also adds  analytical capabilities to FIM. 3D-FIM \cite{katnagallu2022three,vurpillot2007towards,dagan2017automated,klaes2021model}
and aFIM \cite{katnagallu2019imaging,morgado2021revealing} are state-of-the-art techniques that utilize these capabilities to analytically image crystallographic features with atomic resolution in three dimensions besides electron tomography \cite{scott2012electron}. Both 3D-FIM and aFIM are able to provide quasi-analytical images at atomic scale not only for pure materials(Fe,W) but also for complex samples such as FeBSi \cite{klaes2022analytical,de2021evaporation}.
\ignore{
FIM in combination of time of flight mass spectrometry predict the ion trajectories closed to a curved emitting surface modelled at the atomic scale as proposed by \cite{vurpillot2001new,koelling2013direct}.
}
They were recently used to provide insights into crystallographic defects such as vacancies\cite{dagan2017automated,speicher1966observation,park1983quantitative}, dislocations and voids \cite{katnagallu2019imaging,klaes2021development}.
Traditional FIM is also now used for analysing reaction kinetics of various crystallographic facets of Rh for catalytic CO oxidation. \cite{suchorski2020co}. 
Despite the resurgence of FIM based techniques, the quantitative interpretations of imaging contrast enabled by theory have been sparse \cite{Katnagallu2018}. Numerous studies have focused on understanding the image contrast in field ion images
from a theoretical perspective
\cite{FORBES1971, Forbes1972, Forbes1972b, haydock1981field,forbes1985seeing,lam1992field,lam1994theory,forbes2003field,vurpillot2001new}, but a widely applicable framework to predict and interpret
FIM contrast has not yet emerged. 



\begin{figure}
\includegraphics[width=0.5\textwidth]{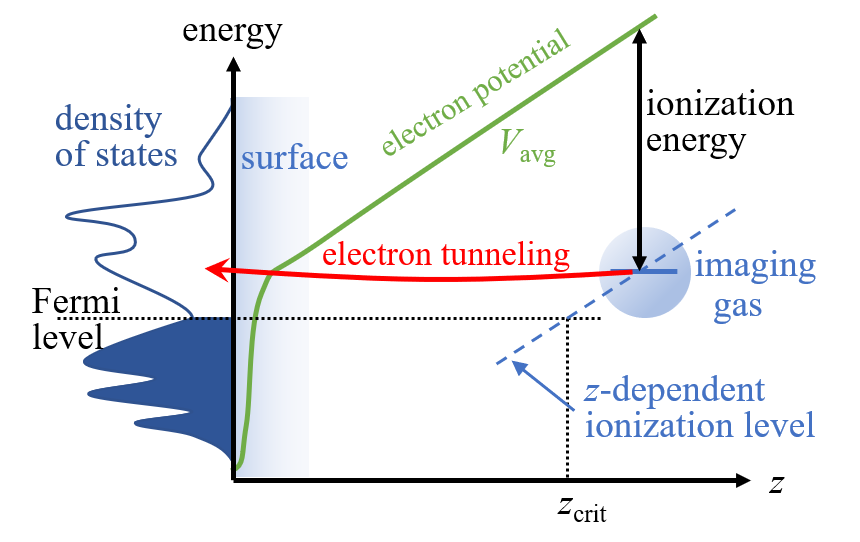}
\caption{Schematic diagram of field ionization. Electrons tunnel from the imaging gas (right) to an energetically aligned empty surface state (left). The plane-averaged electron potential $V_{\rm avg}$ is marked by a green line.  The ionization level (dashed line) and hence the energy of tunneling varies with distance $z$  due to the electric field. For detailed discussion see Eq.~(\ref{eq:energyConserv}). Below the critical distance $z_{\rm crit}$, the ionization level is below the Fermi level and tunneling cannot take place.}
\label{fig:tunneling}
\end{figure}

Under best imaging field conditions, the local imaging contrast is dominated
by the ionization probability at 5-10\,\AA{} above the surface \cite{thesis_Katnagallu2018}.
Ionization near the surface requires an electron transfer
from the gas atom into an empty surface state, see Fig.~\ref{fig:tunneling}. Without
an electrostatic field, this would be energetically impossible
since the ionization level of the imaging gas ($\approx$15-25\,eV) exceeds the work function (\mbox{$\approx$3-6\,eV} for most metals).
The electric field provides the required voltage drop
to overcome this
difference and thereby enables electron tunneling when the
gas atom is beyond a critical distance from the
surface. 
For tunneling into further energetically higher states, the tunneling distance would be accordingly larger. 
\ignore{On the other hand,
tunneling probabilities decay rapidly with increasing
distance (in the presence of a field
even over-exponentially so), limiting the range
of potentially relevant
empty surface states.} 

Unfortunately, previous analytic theories
of field ionization
have focused on simplified scenarios. 
Forbes concluded in his 2003 review of field ionization theory \cite{forbes2003field}:
``A problem is that the complexity of the field-ion
 imaging situation is so great that it is well nigh
 impossible to formulate any valid detailed numerical theory. There
 are image-contrast problems to
 be solved, certainly with some alloy materials. But
 the problem more likely lies in the nature of the
 electronic structure of the substrate, and how this
 influences local fields above individual atoms.''

With the present work, we aim at closing this gap
by a general quantitative framework based 
on density-functional theory (DFT).
For tunneling, we adapt the well-known
the Tersoff-Hamann approach used in STM \cite{tersoff1985theory},
that links tunneling probability to the local
density of states (DOS), which in turn is computed
from the DFT Kohn-Sham wavefunctions.
Preliminary work by us \cite{katnagallu2019imaging,morgado2021revealing}
presented elements of this new framework, but neither provided
a complete, general treatment of the physical
picture nor of the specific computational challenges.

In our previous work \cite{katnagallu2019imaging}, we applied the Tersoff-Hamann approximation in DFT simulations in presence of a field to explain the brighter appearance of Re atoms during FIM imaging of a Ni-Re alloy, focusing on states right above the Fermi level. While such a selective approach explained the chemical contrast in this case, the numerical accuracy in other cases emerged as a critical issue. In our recent work on Ta-doped Ni \cite{morgado2021revealing}, we proposed
that Ta-induced empty states 1-3\,eV above the Fermi
level are responsible for the brightness contrast,
but failed to quantify the interplay of distance
and tunneling energy due to numerical noise of DFT wavefunctions.

\begin{figure}
\includegraphics[width=0.5\textwidth]{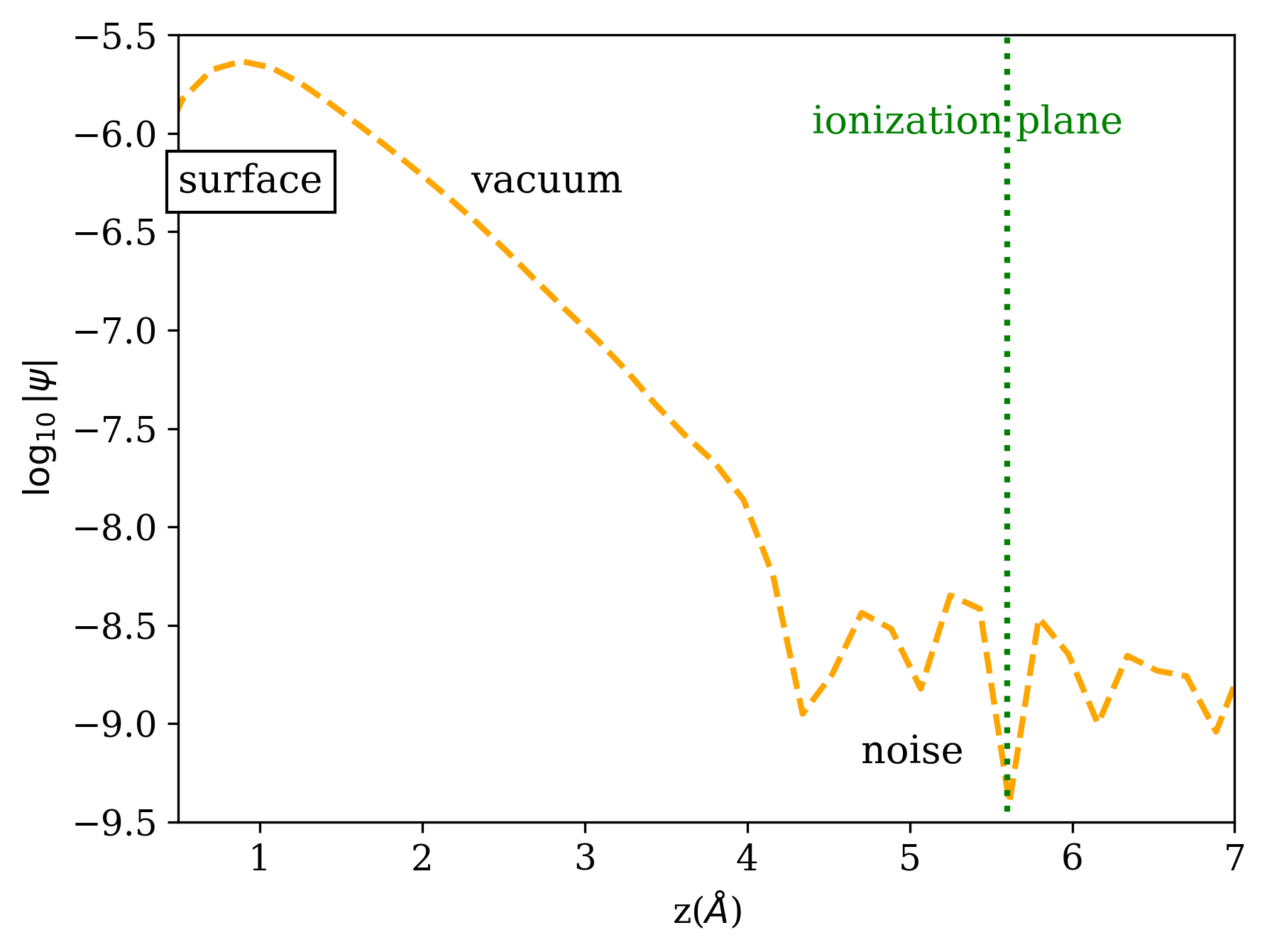}
\caption{Typical evolution of numerical wave functions on top of Ta in a Ni(012) surface in the presence of an electric field as obtained from a standard DFT code. The decay behavior of the wave function (orange curve) is shown for an eigenvalue at the Fermi level. For FIM image contrast, $\psi$ is needed at the ionization plane.}
\label{fig:sketch_idea}
\end{figure}

The computational challenge in these calculations is that the high electric field leads to a very fast decay of wave functions into the vacuum. In many cases, we found that the accuracy of wave functions from standard DFT is insufficient to make quantitative predictions for FIM, as sketched in Fig.~\ref{fig:sketch_idea}.
The standard wave-function optimization algorithms implemented in plane-wave DFT codes are based on the Rayleigh-Ritz method of minimizing the global norm of the residue\cite{payne1992iterative}. This implies that the highest magnitude in the wave function determines the global algorithm's notion of 'large' and 'small'. Hence, they give the best relative accuracy where the wave function amplitude is large. On the downside, in areas where the wave function is small in magnitude, the relative error between the approximate solution and the exact solution can become excessively large even if the total residue is at the numerical limit. The electrostatic field present in FIM further leads to a strong, non-exponential decay of wave functions and they run into a regime where noise dominates. 
Unfortunately, this is exactly the region of space
where the wave functions are needed
to apply the Tersoff-Hamann approximation for FIM.
We considered using an expansion in smooth atom-centered orbitals (or any other type of predefined basis functions “attached” to the surface), but found that this is not an option because the rising, non-trivial vacuum potential at hand induces a decay behavior that deviates from the assumed shape of the basis functions' tails (e.g. Gaussian, exponential).

The solution presented in this work is to recompute the tails of the wave functions with an algorithm that works at the local scale. More precisely, we develop an algorithm that is local with respect to the dominant direction for scale (in the following: $z$), i.e., away from the slab surface\ignore{ to produce an accurate local density of states}.   
For this, we assume that the eigenvalues $\epsilon_i$ and corresponding eigenfunctions $\psi_{i}(\mathbf r)$ close to the slab have been reliably computed. Then, the task is to integrate the underlying second-order differential equation i.e the Kohn-Sham equation from this trusted region along the $z$ direction.

The rest of this article is organized as follows.
In Sec.~\ref{sec:theory_contrast} we explain how DFT in combination
with Tersoff-Hamann tunneling theory enables the prediction
of FIM contrast.
In Sec.~\ref{sec:EXTRA}, we will present our 
tail extrapolation scheme
and show that it is robust even if the wave function amplitude varies over several orders of magnitude along the direction of integration. In Sec.~\ref{sec:results} we apply the new
algorithm to a prototype surface, namely the Ni(012) containing 
substitutional atoms (Ta, W, Re),
and show that we can successfully reproduce the enhanced
brightness observed in experiment
\cite{morgado2021revealing,klaes2022analytical}.

\section{DFT-based theory of FIM contrast}
\label{sec:theory_contrast}
\begin{figure}
    \centering
    \includegraphics[width=.5\textwidth]{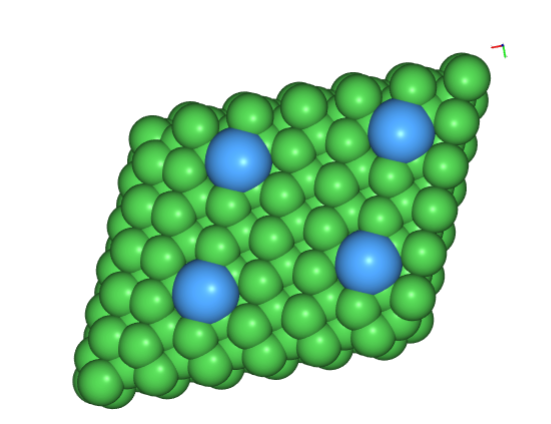}
    
    \caption{Top view of Ta-Ni(012) where Ta atoms (blue) are surrounded by Ni atoms (green). The $3\times3$ surface cell has been repeated in both directions for clarity. }
    \label{fig:structure}
\end{figure}

In the present work, DFT was performed in
the plane-wave PAW formalism with the SPHInX code \cite{boeck2011object} using the Perdew-Burke-Ernzerhof (PBE) exchange-correlation functional \cite{PBE1996} 
The calculation was spin-polarized
with colinear spin and produced a ferromagnetic state. The Ni (012) surface was modeled in the repeated slab approach with 9 atomic layers at the theoretical lattice constant (3.465\,\AA ) and a vacuum separation of 17.5\,\AA.
An electric field of 40\,V/nm 
was included via the generalized dipole correction \cite{genDipole}. 
Metal substitution (Ta, W, Re) at the surface was modeled in a 3$\times$3 surface unit cell, i.e, with 9 surface atoms as shown in Fig.~\ref{fig:structure}.
The 6 topmost layers were relaxed via quasi-Newton optimization
with a redundant internal coordinated Hessian fitted on the fly \cite{ricQN} until the forces were below 0.015\,eV/\AA.
For the structure optimization, an offset $2\times 3\times 1$ 
$\mathbf k$-point sampling was used, equivalent to a 
$\mathbf k$-point spacing of 0.13\,bohr$^{-1}$. For computing
the wavefunctions in the FIM contrast simulation,
the $\mathbf k$-point density in the plane was doubled ($4\times6\times1$).

To simulate the FIM imaging contrast due to the 
electronic structure of the surface, we build on an analogy
to scanning tunneling microscopy (STM).
Both STM and FIM use electron tunneling for forming an image of a solid surface. 
STM relies on electron tunneling between a sample surface and probe tip when a small voltage is applied between them.
In their ground-breaking work \cite{tersoff1985theory},
Tersoff and Hamann found the tunneling current in STM to be proportional
to the surface local density of states (DOS) near the
Fermi level  at the position of the STM tip.
Their derivation equally applies when the spherical tip
is replaced by a single atom.

There are, however, important differences between STM and FIM when
it comes to the
relevant spatial positions, involved states, and the role
of the applied voltage:
In STM, the tip position and the voltage between the tip and the scanned
surface can be controlled. The main role of the voltage is to shift the Fermi level of the tip relative to one of the surface. Explicit field-induced changes are neglected. This is well justified for voltages of a few tenths of
an eV, applied over a distance on the order of 1 nm.
The tunneling current then occurs between states lying between the two Fermi levels. As the tip is retracted from the surface, the tunneling current decays
exponentially.

In contrast, tunneling in FIM occurs between a single electronic
level of the imaging gas (namely the highest occupied one) and a suitably
aligned empty state of the surface, see Fig.~\ref{fig:tunneling}. The externally applied field between the (positively charged) surface and the (negative) counter-electrode far away
makes the energy of the imaging gas orbital rise relative to the surface's electronic structure as the distance to the surface increases, as indicated
by the dashed line in Fig.~\ref{fig:tunneling}.
The further out the atom is, the higher the orbital energy.
Thus, instead of tunneling from a fixed position across a range of states defined
by the applied voltage as in STM,
tunneling in FIM occurs over a range of possible
positions, and in each of these positions to those few surface states
that happen to be energetically degenerate with the ionization level of
the atom at this position. The range of accessible states
is then not defined from the applied voltage. Rather,
high-lying states give negligible contributions because the
tunneling probability rapidly decays as the distance (and thus the
orbital energy) increases, which implicitly imposes an upper limit
of relevance.

Moreover, the field strength in FIM (a few ten V/nm) is higher than in STM, so explicit fields must be considered in simulations \cite{genDipole,katnagallu2019imaging}.
The field not only impacts the charge distribution across the surface
and the relaxation of surface atoms due to Maxwell stress
\cite{genDipole}, but also enhances the wave function decay.
In consequence, the limits of numerical accuracy for
the tails of the wave functions become a critical issue.

Let us now consider how this model of FIM contrast develops into
a computational scheme.
The partial DOS at energy $E$ and position $\mathbf r$ 
is given by
\begin{equation}
\rho(\mathbf r, E) = \sum_{i\sigma\mathbf k} w_{\mathbf k}
|\psi_{i\sigma\mathbf k}(\mathbf r)|^2 
\delta (E-\epsilon_{i\sigma\mathbf k})
\label{eq:pDOS}
\end{equation}
for state index $i$ and spin index $\sigma$. $w_\mathbf k$ denotes
the $\mathbf k$-point weight.
To turn this into a two-dimensional FIM contrast map,
we impose energy conservation for the tunneling process, i.e.,
\begin{equation}
E = \epsilon_{i\sigma\mathbf k} = V_{\rm avg}(x,y,z) - I 
\label{eq:energyConserv}
\;,    
\end{equation}
where $I$ is the (positive) ionization energy of the imaging gas.
$V_{\rm avg}$ is the average potential that the
imaging gas atom experiences. For simplicity we assume here
that $V_{\rm avg}(x,y,z) \approx \overline V(z)$,
the planar average of the potential along the $xy$ plane
parallel to the surface.
This energy conservation condition requires tunneling
into higher-lying states
to occur further away from the surface. Due to the rapid, over-exponential decay of wave functions, the overall contribution
of higher-lying states is effectively dampened. This
relieves us from making \textit{ad hoc} assumptions on which
states are relevant for tunneling. In practice, we truncate
the energy range at $\approx$ 5\,eV above the Fermi level. We
verified that the contribution of the highest of these states to the overall
intensity is negligible.
Combining Eqs.~(\ref{eq:pDOS}) and (\ref{eq:energyConserv}),
the FIM contrast is proportional to
\begin{equation}
F(x,y) = \sum_{i\sigma\mathbf k} w_\mathbf k|\psi_{i\mathbf k}
(x,y,z_{i\sigma\mathbf k,I})|^2
\label{eq:FIM}
\end{equation}
where the sum runs over states above the Fermi level.
The evaluation height $z_{i\sigma\mathbf k,I}$ is implicitly
defined by

\begin{equation}
\overline V(z_{i\sigma\mathbf k,I}) = \epsilon_{i\sigma\mathbf k} + I
\;.
\end{equation}
In practice, we run a search on the discrete $z$ grid, and
then linearly interpolate the DOS between the discrete $z$
points. As the varying contribution of states at different energies
is controlled by their decay behavior, we strongly rely on
an accurate description of the wave function tails.
The tails produced by the standard wave function
optimizers (Rayleigh-Ritz minimization) turn out to be
insufficient, as they are dominated by noise in the region
of interest, i.e., at $z_{i\sigma\mathbf k,I}$.
In the following Section, we therefore present
a new algorithm to recompute these tails, by
numerically integrating the underlying partial differential 
equation in space.

\section{Extrapolation of tail via reverse integration algorithm (EXTRA)}
\label{sec:EXTRA}

\subsection{Reverse Integration}
In DFT, the Kohn-Sham equation  reads\cite{sholl2011density}
\begin{equation}
\left(-\frac{1}{2}\nabla + V_{\rm eff}(\mathbf r)\right)\psi_i(\mathbf r)  = \epsilon_i \psi_i(\mathbf r)
\label{eq:KS}
\end{equation}
using Hartree atomic units ($\hbar$=1, $m_e$=1, $4\pi\epsilon_0$=1). The effective potential is
obtained as
\begin{equation}
V_{\rm eff} = V_{\rm ext}+V_{\rm H}+V_{\rm xc}
\;.
\end{equation}
The external potential $V_{\rm ext}$ defines the
Coulomb interaction between an electron and the collection of
atomic nuclei. The Hartree potential $V_{\rm H}$  describes the classical Coulomb repulsion between 
the electrons, and the exchange-correlation potential
$V_{\rm xc}$ encompasses quantum mechanical
corrections.
In common density-based approximations (e.g. LDA, GGA, meta-GGA), $V_{\rm xc}$ depends on the local density
and vanishes as the density becomes zero. The vacuum potential is therefore dominated by the electrostatic potential $V_{\rm ext}+V_{\rm H}$.

Rewriting Eq.~(\ref{eq:KS}) as
\begin{equation}
\frac{1}{2}\frac{\partial^2}{\partial z^2}\psi_i(\mathbf r) 
=\left\{-\frac{1}{2}\left(\frac{\partial^2}{\partial x^2} +\frac{\partial^2}{\partial y^2}\right)+ V_{\rm eff}(\mathbf r) - \epsilon_i\right\}\psi_i(\mathbf r)
\label{eq:KS_along_z}
\end{equation}

provides the basis to numerically integrate the wave function $\psi_i$ along the $z$-direction. For the sake of readability, we will omit the state index $i$ from the equations in the following.

It is well known from the theory of second-order differential equations in one dimension (1D) that there are two linearly independent solutions.
In our case, a decaying and a rising solution are possible. In forward integration, the decaying solution is always challenging to compute  because numerical noise (e.g. rounding errors due to the finite precision) can produce a small contribution of the rising solution, which then grows as the algorithm proceeds and ultimately dominates. 

To circumvent this problem, one can reverse the direction of integration: by starting deep in the vacuum and integrating towards the surface, the desired solution is growing in magnitude and thus can be easily computed. In consequence, the extrapolation problem is turned into a problem of choosing starting conditions such that the integrated tail solution matches the values of the global optimization near the surface, in our case at the matching plane $z = z_{\rm match}$.

\begin{figure}
    \centering
    \includegraphics[width=0.5\textwidth]{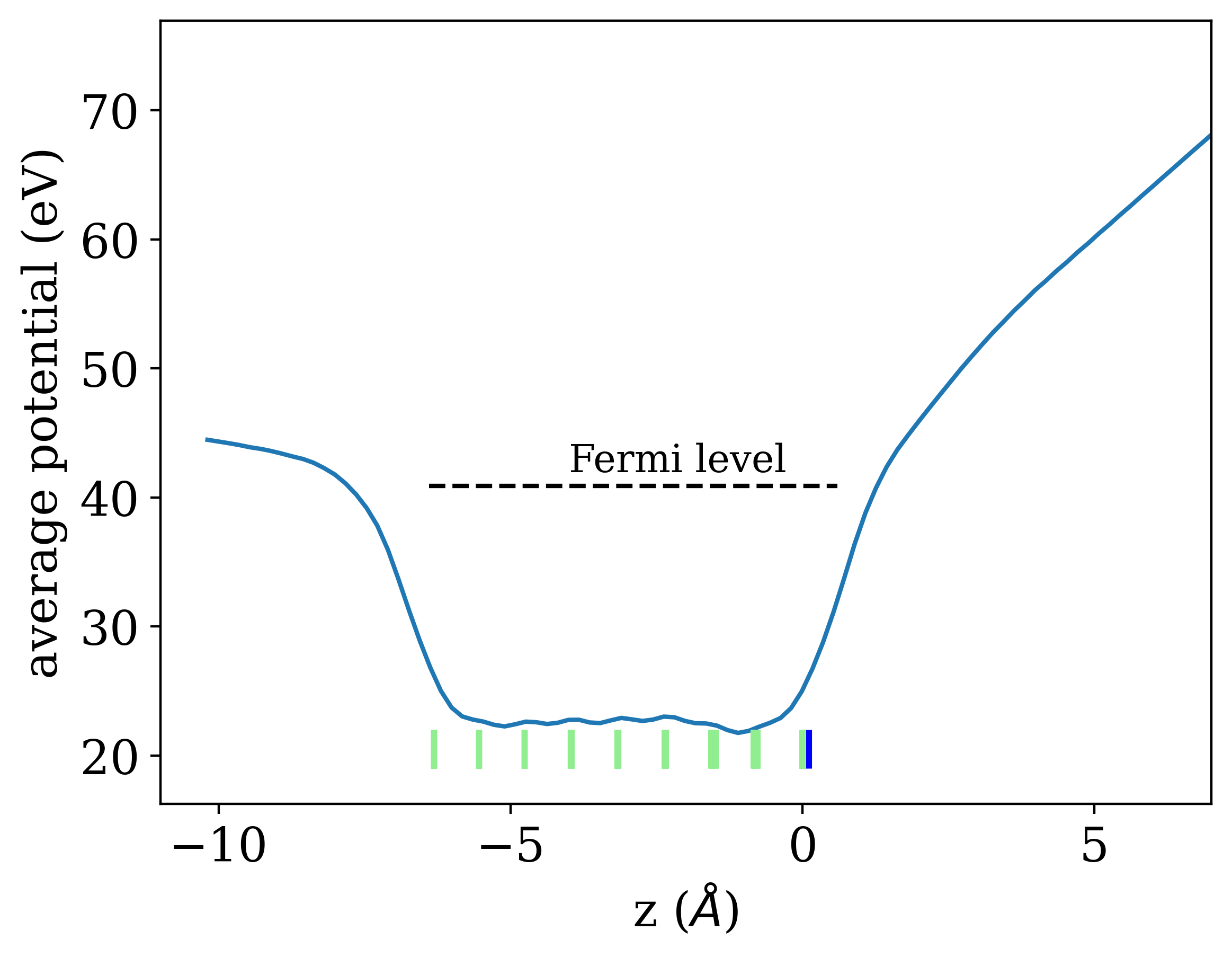}
	 \caption{The effective potential along $z$ averaged over $xy$ plane of a charged Ta-Ni(012) slab with 9 atomic layers as indicated by vertical lines (green: Ni, blue: Ta). An electric field of 40\,V/nm is applied at the top side.}
    \label{fig:average_pot}
\end{figure}

Before we come to the integration algorithm, let us briefly review the key properties of the effective
potential.
In vacuum, where the density becomes negligibly small, the electrostatic part becomes dominant and $V_{xc}$ vanishes. Hence, it is interesting to understand how the electrostatic potential develops far away from the surface.
For this, one can apply a Fourier transform in the $xy$ plane [with wave vector $\mathbf k=(k_x, k_y)$]
to a 'mixed-space' representation $V(\mathbf k, z)$.
As shown in Appendix A, 
the $xy$-averaged electrostatic potential is constant or diverges
linearly along $z$, while the lateral potential variations decay exponentially.
Hence, beyond a certain point, the $xy$-averaged potential will strongly dominate the evolution of the wave functions along $z$. We, therefore, decompose the potential
\begin{equation}
V(x,y,z) = \overline V(z) +\delta V(x,y,z)
\label{eq:full_potential}
\end{equation}
into the average potential $ \overline{V}$ for $k_x=k_y=0$ component and the lateral variation $\delta V$ for
$|\mathbf k|>0$. Fig.~\ref{fig:average_pot} shows the average potential obtained from DFT calculations discussed in Sec.~\ref{sec:results}.

\subsection{Overview of tail extrapolation}
 
\begin{figure}[h]
\centering
\includegraphics[width=0.5\textwidth]{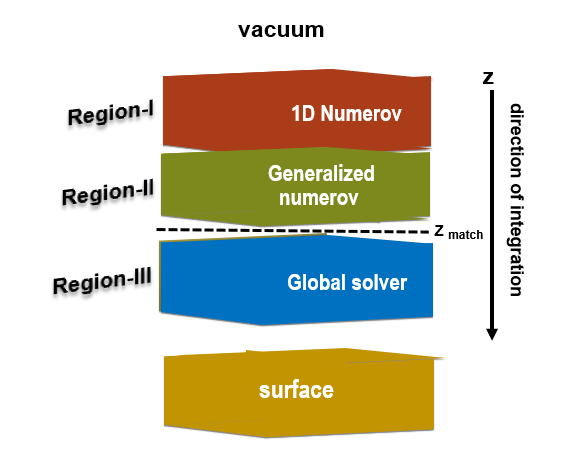}
\caption{Algorithm applied to the various regions above the surface. The reverse integration in $z$ starts from the top to the bottom ($z_{match}$). In region I the 1D Numerov solution is computed. In region II the novel EXTRA algorithm is applied. Region III contains the unmodified plane-wave wave functions obtained from the global solver (plane-wave DFT code).}
\label{fig:region}
\end{figure}

In our algorithm, we split the space above the surface into three
regions, named regions I, II, and III from outside in, as shown schematically in Fig.~\ref{fig:region}. Region III, closest to the surface, is where the
DFT program's global solver provides sufficiently accurate numerical wave functions and hence does not require additional optimization.
For simplicity, we define region III below
a plane parallel to the surface, at height $z_{\rm match}$.
Region I is far away from the slab, where the lateral variations of the potential are negligible. As shown in Sec.~\ref{sec:regionI}, this simplification renders the Kohn-Sham equation separable in the mixed space within region I, and allows us to use 1D Numerov integration along $z$ as an efficient and accurate algorithm to compute wave-function tails. Closer to the slab, in region II, lateral variations in the potential become important and the 1D Numerov would be inaccurate. We therefore generalized the Numerov algorithm to 
three dimensions (see Appendix \ref{app:genNumerov}) to perform integrations
in region II. In Sec.~\ref{sec:regionII} we show that this must be combined
with Fourier-filtering in mixed space to ensure robustness over many orders
of magnitude for the wave-function amplitude. In practice, our Fourier filtering
can be seen as introducing a curved boundary in mixed $(\mathbf k, z)$ space between region I and region II.

The key task to ensure a coherent
wave function across all three regions is to make the 
separately computed wave functions match at
the region I/region II and the region II/region III boundaries, respectively.
For the former, this is readily achieved by rescaling, see Sec.~\ref{sec:regionI} below. For the latter,
where the region III wave functions are authorative, we employ an
iterative procedure summarized in Sec.~\ref{sec:matching}.
As the I/II boundary values serve as starting
values for region II reverse integration towards region III,
we vary these boundary values to minimize
mismatch at the II/III boundary. 
The combined approach, i.e., 1D integration in region I (Sec.~\ref{sec:regionI}), 
Fourier-filtered generalized Numerov integration in region II (Sec.~\ref{sec:regionII}), and the iterative
procedure for determining the I/II boundary values as the key unknowns in
wave-function matching (Sec.~\ref{sec:matching}), is termed EXTRA
 ({\bf EX}trapolation of \textbf Tails via \textbf Reverse integration \textbf Algorithm).

\subsection{Region I: 1D Numerov integration in mixed space}
\label{sec:regionI}

Deep in vacuum, $\delta V$ becomes negligible, and Eq.~(\ref{eq:full_potential}) reduces to
\begin{equation}
V(x,y,z) \approx \overline V(z)
\;.
\end{equation}


Within this approximation, an in-plane Fourier transform makes the
Kohn-Sham equation, Eq.~(\ref{eq:KS_along_z}), separable in $\mathbf{k}$ and z, i.e., in mixed space
\begin{equation}
\frac{1}{2} {\partial^2 \psi(\mathbf k,z) \over \partial z^2} =  \left\{\frac{1}{2} |\mathbf k|^2 + \overline V(z)  - \epsilon\right\}\psi(\mathbf k,z)
    \label{equation:mixed_space}
\end{equation}
with both growing and decaying solutions if $\epsilon < \overline V(z)$ for all $z$. For recomputing the decaying tails, 1D Numerov is a feasible algorithm \cite{purevkhuu2021one,bennett2015numerical}. For this, one performs reverse Numerov integration, starting deep in the vacuum with arbitrary non-vanishing initial values, and rescales the intermediate solution such that it matches a given value $\psi(\mathbf k, z_{\rm start})$ at the boundary $z_{\rm start}$ between regions I and II.

The Numerov method is a finite difference method that calculates the shape of the wave function by integrating step-by-step along a grid. The one dimensional Schr\"odinger equation is solved using Numerov algorithm\cite{kenhere2007bound} in form of
\begin{equation}
  \psi_{n+1} = \frac{2(1-\frac{5}{12}h^2k^2_n)\psi_n- (1+
  \frac{1}{12}h^2k^2_{n-1})\psi_{n-1}}{1+\frac{1}{12}h^2k^2_{n+1}}
\end{equation}
where
\begin{equation}
k_n^2 = 2\epsilon-2\overline V(z_n)-|\mathbf k|^2
\;.
\end{equation}

There are different variants of Numerov \cite{simos2009new,pillai2012matrix} developed in the past for the approximate solution of Schr\"odinger equation.
The discretization error of the Numerov algorithm is $\mathcal O(h^6)$.

The numerical accuracy of the Numerov algorithm  for rising solutions over many orders of magnitude arises
from the locality: only the
previous two values are needed, and these
lie at a very similar magnitude.
Uncorrelated numerical round-off errors cannot grow faster
than the exact solution, because they either are proportional to the desired
solution or belong to the linearly
independent, decaying solution. This also
ensures that starting from arbitrary
values will always converge towards the
rising solution after a warmup phase.

\subsection{Region II: Fourier-filtered generalized Numerov integration}
\label{sec:regionII}
\subsubsection{Generalized Numerov algorithm}
\label{sec:genNumerov}
In order to numerically integrate the 3D Schr\"odinger equation along the $z$ direction, we propose a generalized Numerov algorithm given in Appendix~\ref{app:genNumerov}. The working equation is
[cf. Eq.~(\ref{eq:gen_Numerov_final})]
\begin{eqnarray}
&&\left[1+\frac{1}{6}\zstepn^2\left\{\epsilon-\hat{V}(z_{n+1})\right\}\right]\psi_{n+1}
\nonumber\\&=& 2\left[1-\frac{5}{6}\zstepn^2\left\{\epsilon-\hat{V}(z_{n})\right\}\right]\psi_n
\nonumber\\&&
-\left[1+\frac{1}{6}\zstepn^2\left\{\epsilon-\hat{V}(z_{n-1})\right\}\right]\psi_{n-1}
\label{eq:gen_Numerov_text}
\end{eqnarray}

Eq.~(\ref{eq:gen_Numerov_text}) represents an in-plane partial differential equation [with linear operator $\hat V(z)$]
for $\psi_{n+1}$ with
a known right-hand side that depends on the values for the two previous steps $z_n$ and $z_{n-1}$. By solving this differential equation numerically using a standard (plane-global) iterative algorithm, one can step-wise proceed along the $z$ direction.
The in-plane kinetic operator
is computed in mixed space, while
the potential is applied in real space,
using fast Fourier transforms to switch
between these two spaces.
To solve the discretized differential equation,
we employ the root finding solver of
the Scipy optimization module\cite{virtanen2020scipy} (scipy.optimize.root) with
Krylov subspace iterations
\cite{kelley1995iterative} and a
numerical approximate inverse Jacobian.

The key advantage of this procedure is that the numerical solver of the in-plane equation
must only deal with scale variations within the plane, while the huge changes in magnitude
along the $z$ direction are taken care of by the explicit 
iteration $(\psi_{n-1}, \psi_n)\rightarrow \psi_{n+1}$.

\subsubsection{High-frequency noise issues with unfiltered generalized Numerov}
We have tested the Generalized Numerov numerical integration in the reverse direction in a region near the surface where the global solver provides a nicely decaying reference solution discussed in the 
Supplementary material.

Generalized Numerov is stable against numerical noise from the unwanted solution (i.e., the exponential rising solution into vacuum). Unfortunately, when integrating towards the surface, it produces a rapid increase of contributions from high-frequency Fourier components in the plane that are absent from the reference solution. This is not a failure in principle, as we can expect from the separable approximation, Eq.~(\ref{eq:1D_seperable}) that in-plane frequency coefficients for high $\mathbf k$ values have steeper slopes along $z$ compared to the low-frequency ones as shown in Fig.~\ref{fig:decaying}. Yet, this makes the Generalized
Numerov algorithm of Appendix~\ref{app:genNumerov} 
sensitive to high-frequency noise.

\begin{figure}
     \centering
     \includegraphics[width=0.5\textwidth]{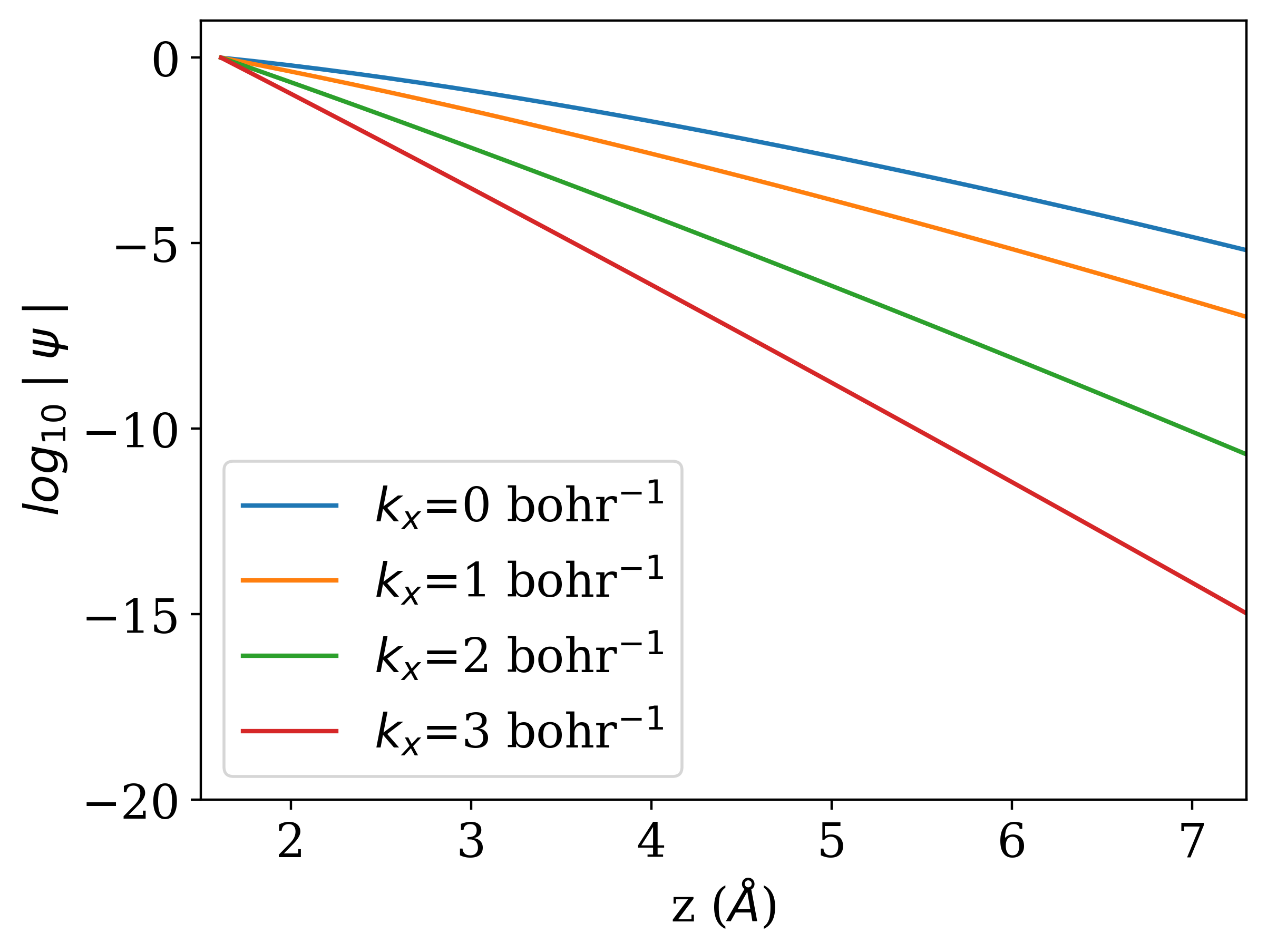}
     \caption{1D Numerov wave-function magnitude as a function of surface-ion distance $z$ using the potential shown in Fig.~\ref{fig:average_pot}. The topmost nuclear positions are located at $z$=0. The decay rate depends on in-plane Fourier components $k_x$. The wave functions are calculated for an eigenvalue at the Fermi level and normalized with respect to $z_{\rm match}$=1.7\,\AA. }
     \label{fig:decaying}
 \end{figure}
 
\begin{figure}
    \includegraphics[width=.5\textwidth]{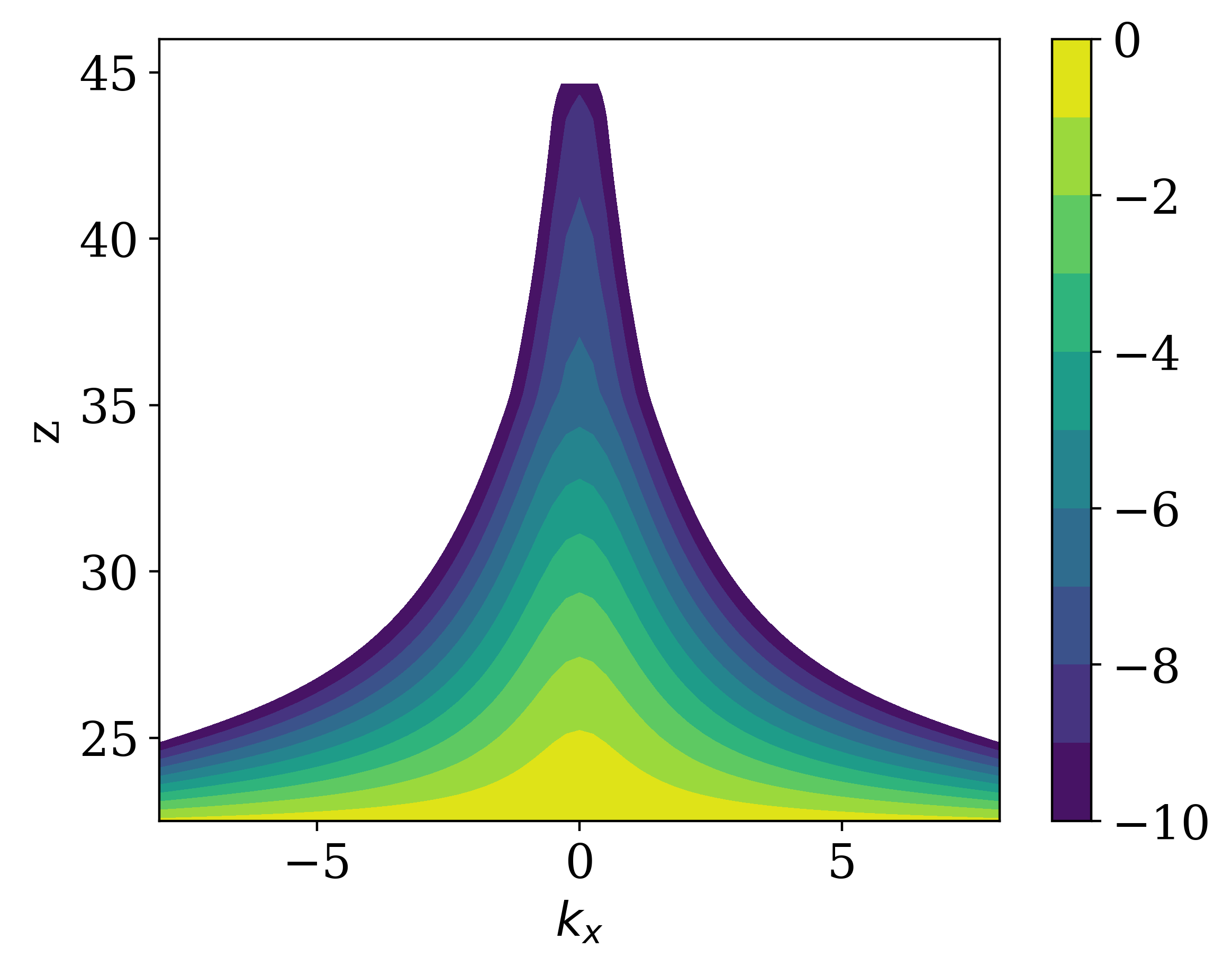}
    \caption{Iso-magnitude contours of wavefunctions above Ta-Ni(012) plotted in mixed space. The wave functions were generated from 1D Numerov using the average potential (see Fig.~\ref{fig:average_pot}), and were normalized with respect to 
    $z_{\rm match}$=22\,bohr.}
    \label{fig:iso_contour}
\end{figure}
The reason behind the discrepancy is visualized in  Fig.~\ref{fig:iso_contour} which illustrates how the wave-function magnitude develops for different $k_x$ values.

As shown in Fig.~\ref{fig:iso_contour}, iso-magnitude contours in mixed $\mathbf (k_x,k_y,z)$ space are not parallel to the $\mathbf (k_x,k_y)$ plane. In consequence, we find that it is not sufficient to find an algorithm that has a local scale in only $z$, but actually, one that respects the local scale in $\mathbf (k_x,k_y,z)$.
To circumvent the above issue, we additionally employ a $z$-dependent Fourier filtering in the $xy$ plane. The challenge is to distinguish between noise and the true signal. Fortunately, for each mixed-space coefficient, we can estimate the expected magnitude relative to the matching plane  from the 1D separable equation, Eq.~(\ref{equation:mixed_space}). We can use this to make a Fourier-filtering at "equal magnitude".

\subsubsection{The iso-magnitude boundary for regions I/II}
\label{sec:isomagnitude}
At the boundary between region II/region III high frequency Fourier components obtained by the Generalized Numerov algorithm grow rapidly and need to be filtered out. To do this we make the boundary k-dependent using the iso-magnitude condition 
\begin{equation}
\psi_{1d}(\mathbf k,z_{\rm start}(\mathbf k))/\psi_{1d}(\mathbf k,z_{\rm match}) = \eta
\label{eq:iso_magnitude}
\end{equation}
where $\eta$  defines the magnitude threshold. Eq.~(\ref{eq:iso_magnitude}) defines a finite $z_{\rm start}(\mathbf k)$ beyond which the coefficients can be effectively ignored (set to zero) for the generalized Numerov step as shown in Fig.~\ref{fig:iso_contour}.
The choice of $\eta$ is not overly problematic
in practice. We have successfully employed values of $10^{-6}$, $10^{-8}$, and even $10^{-20}$ and observed negligible
differences between the results. If $\eta$ is chosen
too small,
high-frequency noise occurs. If chosen too large,
the original DFT wave functions are not well reproduced
near the matching plane (where the intrinsic
noise is small).

The values $\psi(\mathbf k,z_{\rm start}(\mathbf k))=\psi_{\rm start}(\mathbf k)$
are used as the initial conditions for the generalized Numerov integration. They
fully determine the shape of the wave function inside region II.
For initializing the previous value we use 1D Numerov to estimate $\psi(\mathbf k,z_{\rm start}(\mathbf k)-\zstep)$. Similarly, we use 1D Numerov to extend $\psi$ beyond the filtering boundary in an approximate way namely ignoring the effect of in-plane scattering due to the in-plane potential variant ions $\delta V$.

 In this way, the iso-magnitude contour is effectively treated as our dividing boundary between region I and region II in mixed space.
  At this boundary, we initialize the Fourier components for
 the region II integration. In short only coefficients inside the boundary are included in the generalized Numerov for region II.
 Outside this contour boundary, i.e., in region I, we rescale the 
 precomputed 1D Numerov solutions to match the boundary value.
 
 We note in passing that we can combine the iso-magnitude boundary
 condition with a maximum for $z_{\rm start}$ based on the in-plane
 lateral variations $\delta V$. In such a case, we cap
 the contour when the lateral variations become negligible,
 and thus the 1D Numerov integration is accurate (and far more efficient).

\subsubsection{Fourier filtered generalized Numerov}

To summarize the Fourier-filtered generalized Numerov, the
iteration proceeds as follows
\begin{enumerate}
    \item Given $\psi(\mathbf k, z_{n-1})$ in mixed space and
    $\psi(x,y,z_n)$ in real space 
    on a regular discretization grid,  Fourier transforms the
    latter one to mixed space via Fast Fourier transforms (FFT).
    \item Set $\psi(\mathbf k, z_n) := 0$
    where $z_n > z_{\rm start}(\mathbf k)$.
    \item For cases at the boundary, where $z_n = z_{\rm start}$,
          set
          \begin{eqnarray*}
          \psi(\mathbf k, z_n) &=& \psi_{\rm start}(\mathbf k)\\
          \psi(\mathbf k, z_{n-1}) &=&
          \psi_{\rm start}(\mathbf k)\cdot
          \frac{\psi_{1d}(\mathbf k,z_{n-1})}{\psi_{1d}(\mathbf k,z_{n})}
          \end{eqnarray*}
    \item Save $\psi(\mathbf k, z_n)$ for the next iteration
    \item Fourier transform both $\psi(\mathbf k, z_{n-1})$ 
          and $\psi(\mathbf k, z_n)$ to real space
    \item Perform generalized Numerov propagation by solving
          Eq.~(\ref{eq:gen_Numerov_text}), yielding the real-space
          for the next iteration $z_{n+1}$.
\end{enumerate}
When $z_{\rm match}$ has been reached, the missing
region I tails 
 are added to the mixed-space representation 
by setting
\begin{equation}
\psi(\mathbf k, z_{n}) =
          \psi_{\rm start}(\mathbf k)\cdot
          \frac{\psi_{1d}(\mathbf k,z_{n})}{\psi_{1d}(\mathbf k,z_{\rm start}(\mathbf k))}
\end{equation}
for all $z_n > z_{\rm start}(\mathbf k)$. Afterward,
the real-space representation can be recomputed from
this via in-plane FFTs.

\begin{figure}
    \centering
    \includegraphics[width=.5\textwidth]{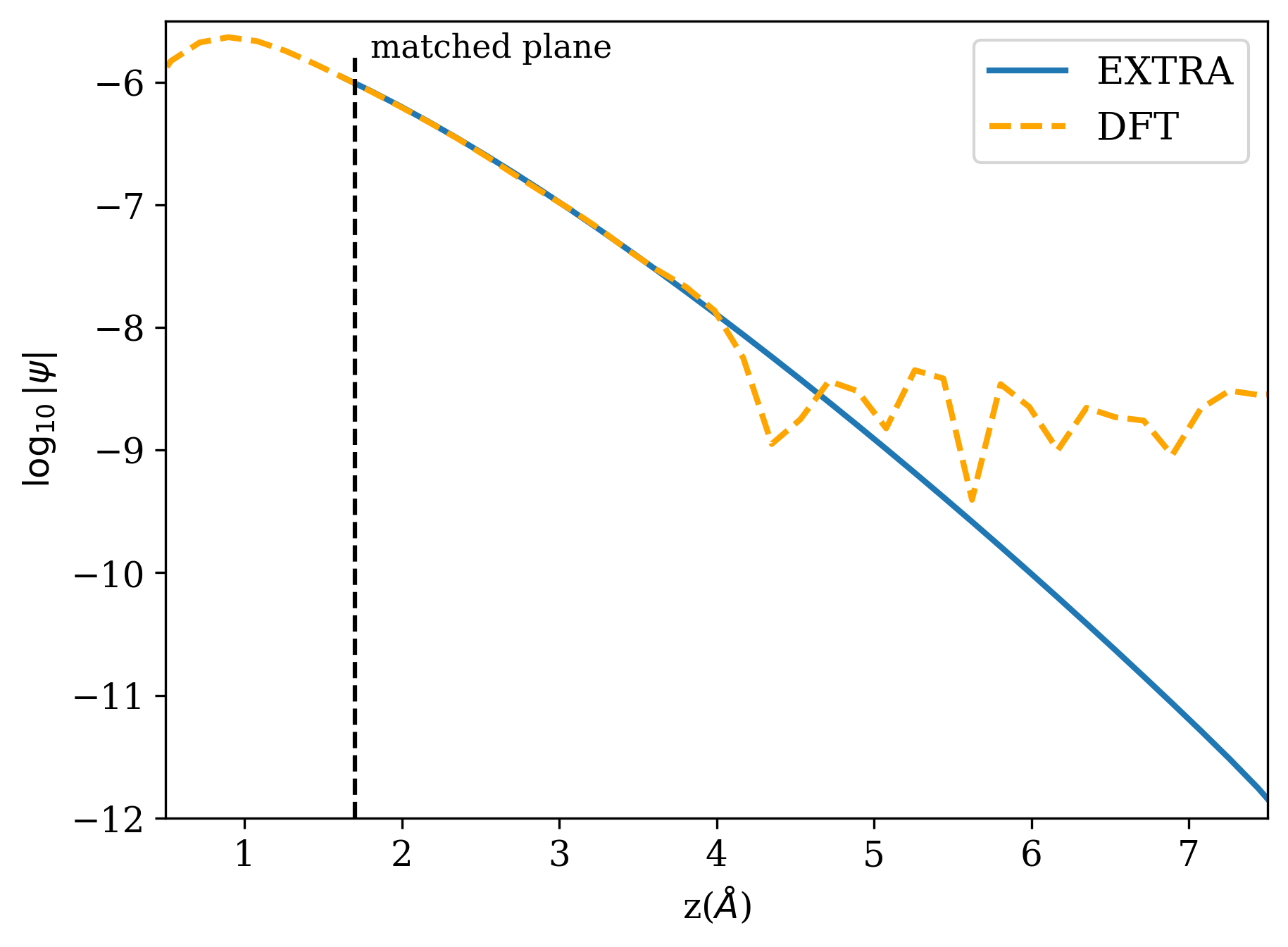}
    
    \caption{Comparison of EXTRA and DFT computed wave-functions of Ta-Ni(012) along the direction $z$. The topmost nuclear positions are located at $z$=0. The wave functions correspond to an eigenvalue at the Fermi level. The position of the matched plane (dashed line) is located at $z_{\rm match}$ = 1.7\,\AA{}  where both curves have been aligned.}
    \label{fig:EXTRA_global}
\end{figure}

\subsection{Iterative determination of $\psi_{\rm start}$}
\label{sec:matching}

The final step is to determine the starting values 
$\psi_{\rm start}(\mathbf k)$ such that integration across
region II according to the algorithm described above yields values $\psi_{II}(x,y,z_{\rm match})$
at the matching plane
that agree with the desired values. 
These values are given by the original DFT
wave function in region III, i.e., by $\psi_{III}(x,y,z_{\rm match})$.

For this purpose, we define the residue
\begin{equation}
 R(\mathbf k) = \psi_{III}(\mathbf k,z_{\rm start}) -  \psi_{II}(\mathbf k,z_{\rm match})
\label{eq:matching}
 \end{equation}
 that implicitly depends on $\psi_{\rm start}(\mathbf k)$,
and solve the multidimensional root-finding problem $R = 0$,
treating $R$ as a function of $\psi_{\rm start}$.
However, there is a huge difference in magnitude between the residue $R$
(which is similar in scale to the wave function at the matching plane),
and the starting values $\psi_{\rm start}$ at the iso-contour boundary. The latter are smaller by a factor $\eta$, cf. Eq.~(\ref{eq:iso_magnitude}).
To accommodate the scale,
we iterate on
\begin{equation}
\psi_{\rm init}(\mathbf k) = \psi(\mathbf k)\cdot
\frac{\psi_{1d}(\mathbf k,z_{\rm match})}{\psi_{1d}(\mathbf k,z_{\rm start}(\mathbf k))}
\;.
\end{equation}
$\psi_{\rm init}$ can be thought of as the boundary values 
rescaled to the matching plane via the 1D Numerov approximation.
We use this flexible definition
rather than the constant $\eta$ to accommodate situations in
which we limit $z_{\rm start}(\mathbf k)$ to a maximum 
based on the magnitude of $\delta V$, as explained in Sec.~\ref{sec:isomagnitude}.
The root finding algorithm, $scipy.optimize.root$
with Krylov iteration and numerical inverse Jacobian estimation \cite{kelley1995iterative}, is then used to solve for the 
starting values, completely analogous to our solution
of the generalized Numerov propagation, Eq.~(\ref{eq:gen_Numerov_text}),
see Sec.~\ref{sec:genNumerov}. Fig.~\ref{fig:EXTRA_global} illustrates the comparison of the original, noisy wave function from the global DFT solver with one from EXTRA on the log scale. It demonstrates that EXTRA overcomes the limitations of the global solution.

\section {Investigation of substitutional impurities in Ni}
\label{sec:results}
\subsubsection{Chemical contrast for Ta in Ni(012)}
In this Section, we will illustrate that the EXTRA algorithm
allows us to overcome the accuracy limitations that prevented
direct simulations of FIM contrast. For this evaluation, we choose the
case of substitutional Ta in Ni, before we analyse
systematic trends for 5d elements series (Ta, W, Re) in
Sec.~\ref{sec:TaWRe}. We have selected these systems because transition metal solutes in Ni have been demonstrated to enhance high-temperature deformation resistance, a critical property for Ni-based superalloys \cite{URREHMAN201754}.
In a recent aFIM study with Ne as an imaging gas,
Morgado \textit{et al.} investigated
segregation in Ni alloys with 2\% Ta \cite{morgado2021revealing}.
They observed that Ta
was imaged in FIM more brightly than Ni. 
This finding was 
qualitatively explained by DFT calculations performed by 
some of the present authors. The DFT calculations showed that Ta-related states appear energetically at 1-3 eV above the Fermi level, while only a few Ni states in the spin
minority channel are available for
tunneling electrons up to 1\,eV above the Fermi level.
However, due to the accuracy limitations, we could not actually compute the FIM contrast at relevant
ionization energies, nor verify that Ta-related states at higher
energies give at all a brighter signal than the lower-lying Ni
states.
More recently, Klaes \textit{et al.} \cite{klaes2022analytical} provided 
quantitative data for the spot intensity distribution of the Ni-Ta alloy in FIM.
They show two distinct maxima in the intensity histogram of imaged atoms (Ta and Ni) from field ion images, and
concluded that intensity can be used to deduce the chemical identity of the imaged
atoms with some confidence. We will compare our computed intensity ratios
with this experimental data in Sec.~\ref{sec:comparison with experimental data}

\begin{figure*}
     \centering

     \hfill
     \hfill
         \includegraphics[width=0.44\textwidth]{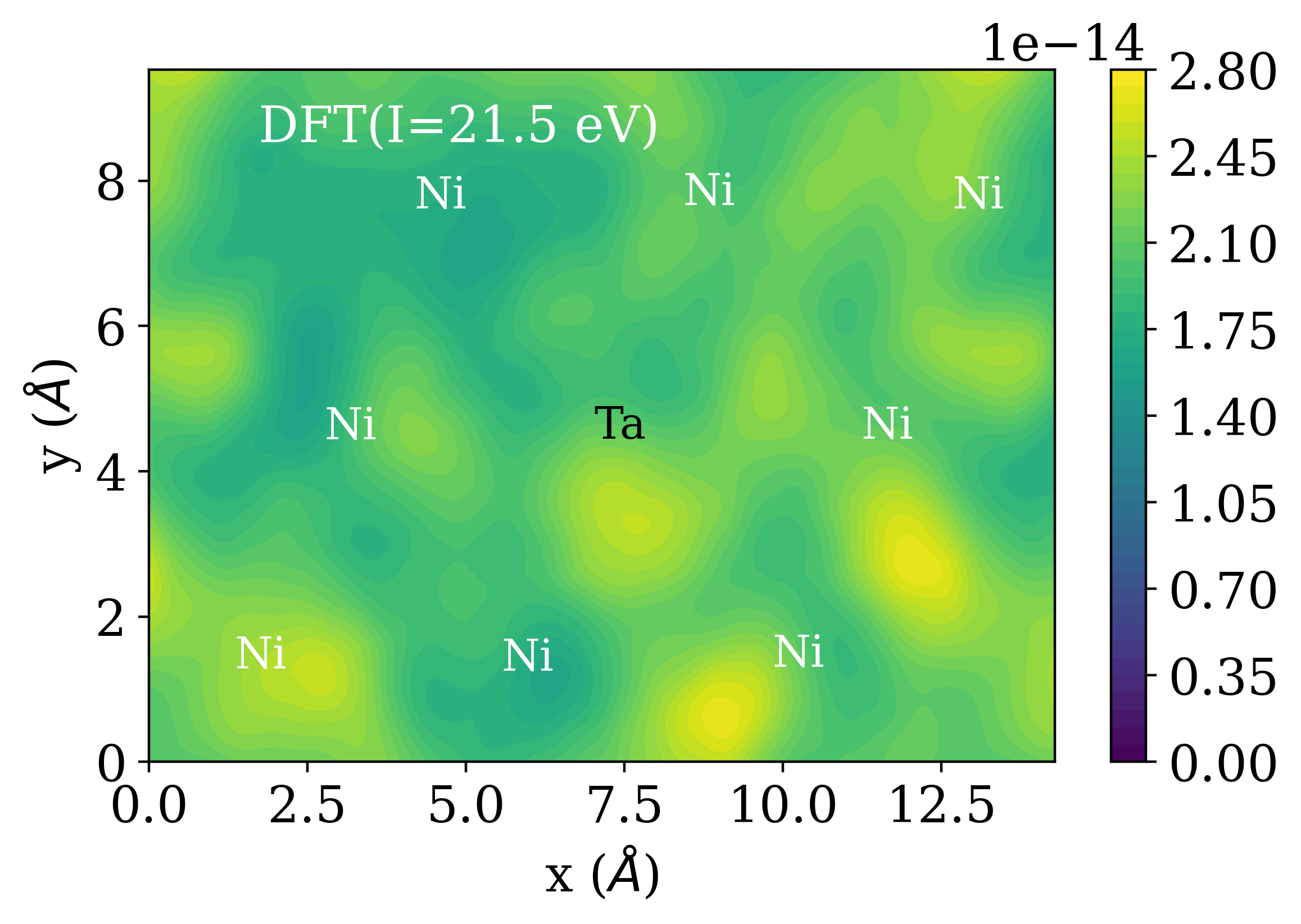}
     \hfill
     \hfill
     \hfill
     \includegraphics[width=0.44\textwidth]{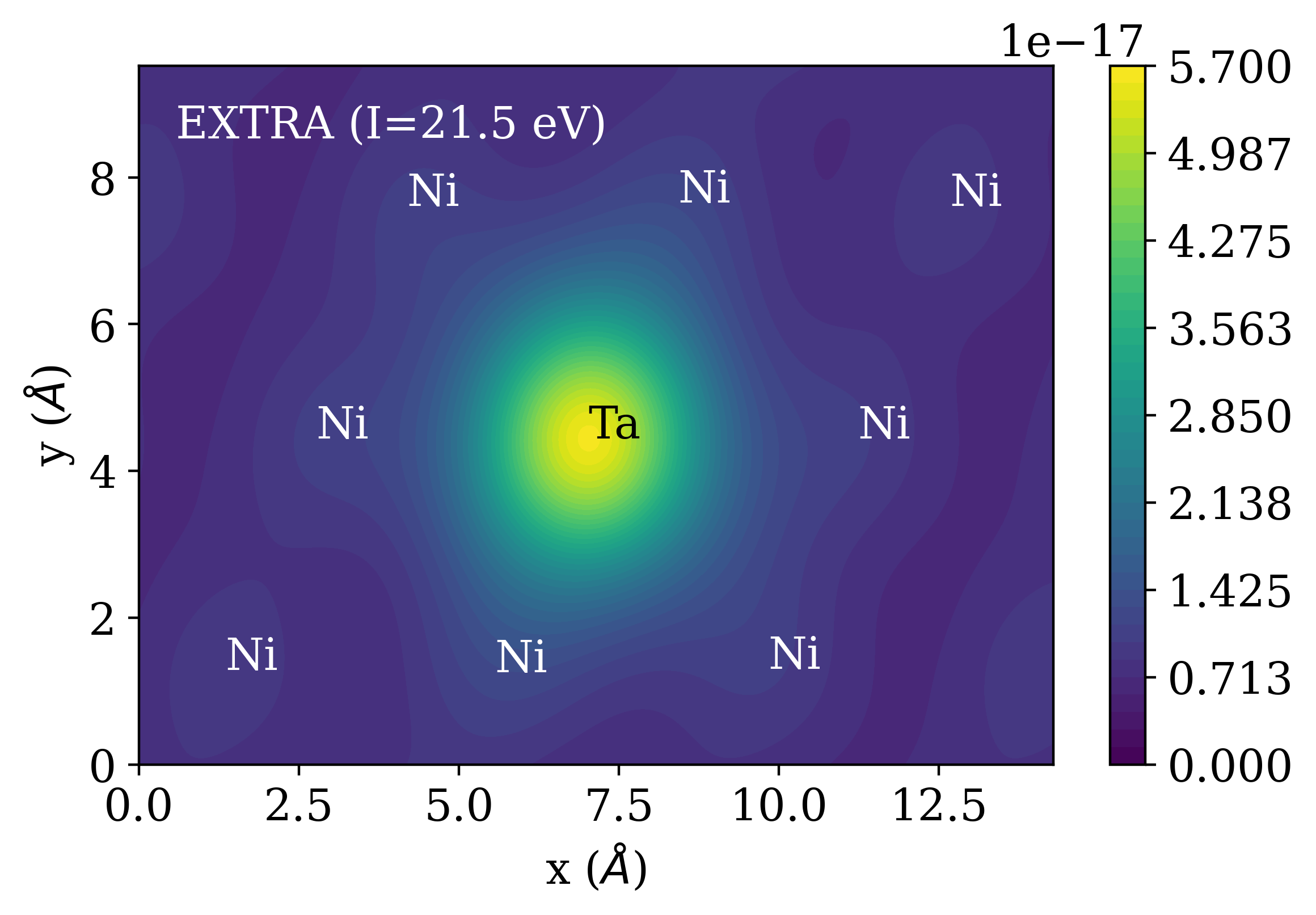}
        \caption{Simulated FIM images of Ta-Ni(012) for an ionization energy of 21.5 eV. Left: partial DOS obtained from the original (noisy) DFT wave functions. Right: refined results from EXTRA. In-plane positions of the top layer atoms are indicated in the graph. }
        \label{fig:Fim_Ta}
\end{figure*}

\begin{figure}
    \includegraphics[width=0.44\textwidth]{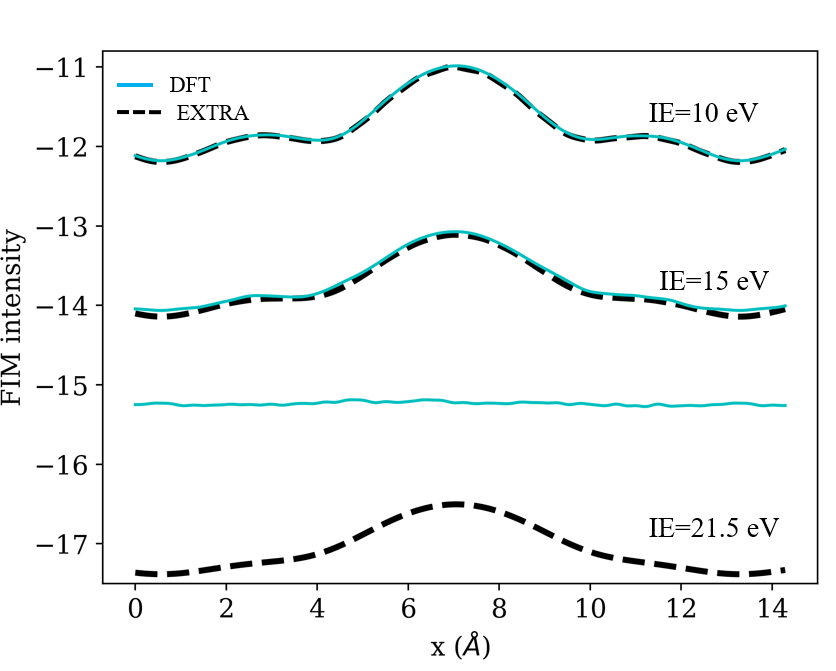}
     \caption{1D line scans of the simulated FIM intensity from Fig.~\ref{fig:Fim_Ta} on a log scale. The line runs along the $x$ axis across the Ta position. }
     \label{fig:1Ddft_extra}
\end{figure}

\ignore{ Note that increasing $I$ by 5\,eV shifts the evaluation region by 2.4\,bohr in a field of 40 V/nm.}
Fig.~\ref{fig:Fim_Ta} illustrates the results at a realistic ionization energy of 21.5eV. In the left plot, we show simulated FIM Image from the original DFT wave functions. At ionization energy of 21.5\,eV, the simulated
contrast contains only noise. On the other hand, the results obtained
with EXTRA in the right plot show a very clear contrast between Ta and surrounding Ni atoms. As a result, the observed brightness contrast of Ta in Ni shows the effectiveness of our novel approach, EXTRA.

What happens to noise at a distance closer to the surface? As described in Section~\ref{sec:theory_contrast}, we can treat the ionization energy $I$ as a tunable parameter to evaluate the FIM contrast at different heights. At lower ionization energies of 10\,eV and 15\,eV, there
is a single bright spot arising from Ta even from the raw DFT wavefunctions (Supplementary material).
\ignore{At low ionization energies, there are also weak signals
from the other surface atoms (Ni), but these are absent
at higher energies.} 
Fig.~\ref{fig:1Ddft_extra} shows line profiles of the simulated FIM intensity at different ionization energies. The line scans run along the $x$ axis across the Ta atom and its two Ni neighbours. The solid lines illustrate the DFT results while EXTRA results are shown by dashed lines. At I = 10\,eV and 15\,eV i.e more closer to the surface, one can not see noise. However, the relative contrast clearly changes at 10\,eV and 15\,eV.  It is concluded that peak shape varies with ionization energy. This highlights that one has to simulate using the correct ionization energy to do a quantitive comparison with the experiment.
The comparison of different ionization energies shows that all features become broader with increasing ionization energy (and hence increasing distance from the surface). One can clearly see the intensity difference of five orders in magnitude at I = 10\,eV and 21.5\,eV, that clearly indicates the stable performance of our algorithm EXTRA over several orders of magnitude. At higher ionization energy of 21.5\,eV the noise level from DFT exceeds by two orders of magnitude the expected true signal from EXTRA. The simulated FIM contrast is hence strongly improved by the extrapolation of wave-function tails. 

\ignore{We note that this finding is not observed in the experiment, which clearly images Ni atoms
even near Ta atoms. The reason for this strong exaggeration
of brightness contrast is not entirely clear. We suspect that an adsorbed imaging gas layer present in the experiment
may enhance spatial resolution. We will investigate this
in more detail in a forthcoming publication.} 

\subsubsection{Comparison to experiment}
\label{sec:comparison with experimental data}
Klaes \textit{et al.} \cite{klaes2022analytical} studied a NiTa alloy using 3D-FIM. Using two-pass algorithm \cite{klaes2021development} they have analysed the intensity spectrum of imaged atoms right before evaporation. Based on their observations, we performed a comparison by calculating the peak intensity of Ni and Ta. Referring to Fig. 4b of Klaes et al. \cite{klaes2022analytical}, we calculate the peak intensity ratio from intensity distribution. The intensity spectrum runs between 0 and 250 for the corresponding Ni and Ta spots. Ni appears as a broad feature around 44 (FWHM: $\approx$50) while Ta feature peaks at 222 (FWHM: $\approx$20). Considering these two different intensities we compute the
experimental ratio as $I_{Ta}/I_{Ni}$=5.05. 
In our current work, we compare intensity from pure Ni metal and added impurities. From FIM simulation we found the peak intensity ratio to be $I_{Ta}/I_{Ni}$ = 6.18. Our theoretical ratio is found to be in reasonable agreement with experimental value, being   larger than the experimental observation by 23\%. 
\ignore{We also predict the full width at half maximum (FWHM) of spectrum from cited experimental work. For the case of Ni it is found to be 51\% of top site of low intensity peak, for Ta it is 20\% of top site of high intensity peak. This confirms that it is possible to differentiate chemical identities of atoms.}


\subsubsection{Effect of impurities on contrast}
\label{sec:TaWRe}
\begin{figure}
     \centering
     \includegraphics[width=0.44\textwidth]{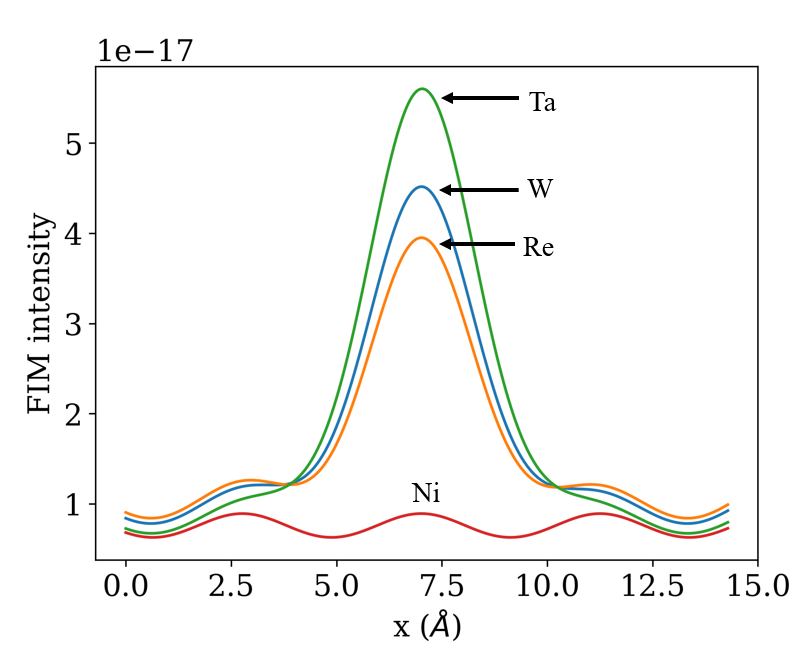}
     \caption{1D line scans of the simulated FIM intensity for Ta, W, and Re on Ni(012) running along the $x$ axis (horizontally see Fig.~\ref{fig:Fim_Ta}) across the central atom. The red curve shows the  FIM intensity for pure Ni. }
     \label{fig:peak_intensity}
\end{figure}

\begin{figure}
    \centering
        \includegraphics[width=0.44\textwidth]{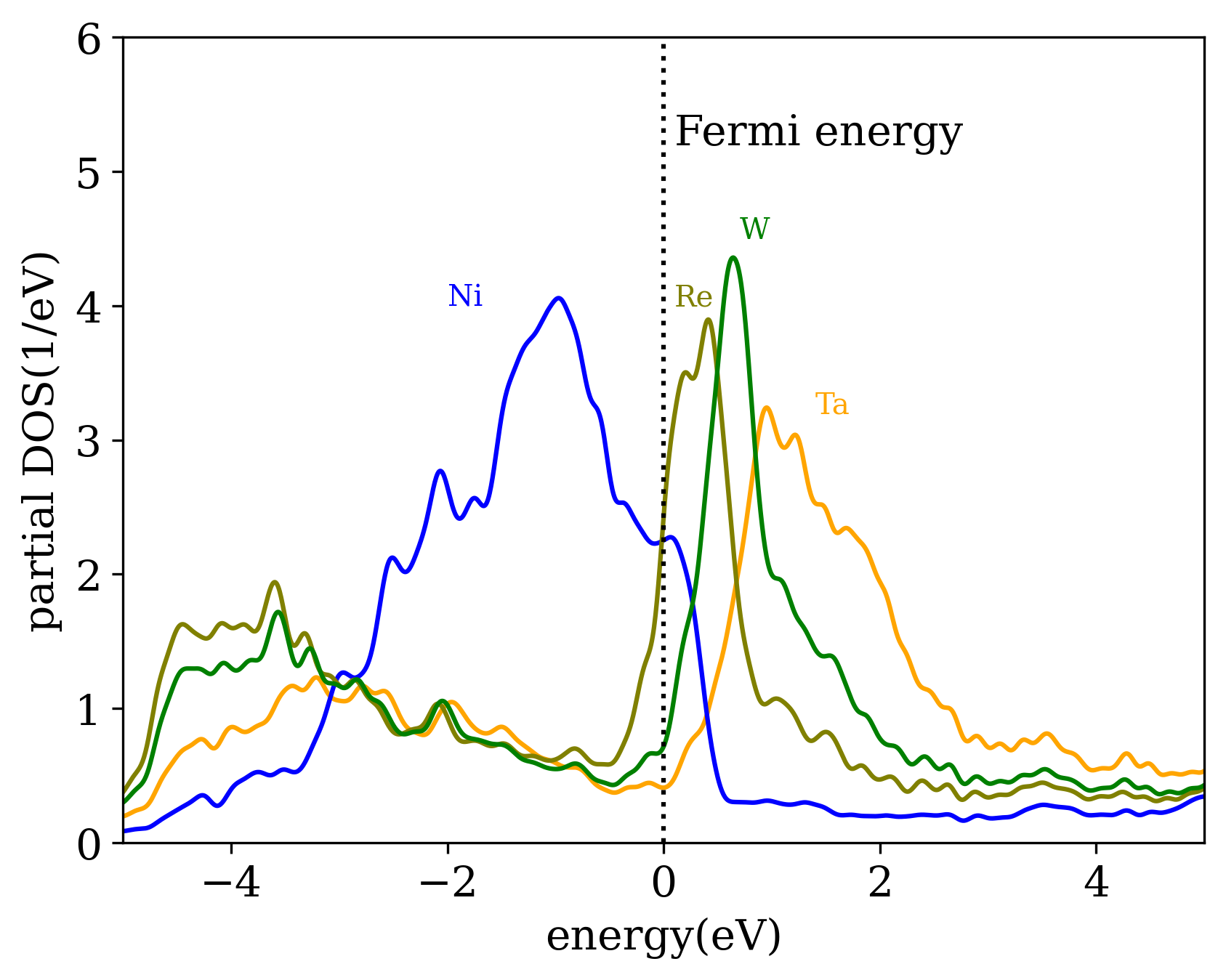}
        
            \caption{Atom-projected density of states for W, Ta, and Re atoms located in the top surface layer of Ni(012). Fermi energy is referred at zero.}
        \label{fig:partial_dos}
\end{figure}

Following the same approach, we have investigated the FIM appearance of Tungsten (W) and Rhenium (Re) on Ni(012) surface. The FIM contrast maps are similar to the Ta case
(Fig.~\ref{fig:Fim_Ta})
with a single bright spot from the substituted atoms (see Supplementary material).
However, the brightness enhancement varies with the chemical element. To illustrate
this, Fig.~\ref{fig:peak_intensity} provides the FIM intensity profiles for Ta, W,
and Re substitution along a
line running through the central atom and its Ni neighbours along the $x$ direction.
For reference, we included the case of a pure Ni surface.
The comparison shows that the maximum enhancement in contrast is for the Ta atom and Re shows the minimum i.e Ta$>$W$>$Re. 
From the line profile plots,
we can extract the relative peak intensities of these elements. For the case of Tungsten it is $I_{W}/I_{Ni}$ = 5.06 and for Re $I_{Re}/I_{Ni}$ = 4.38. 

The enhanced brightness of 5d elements emerges from the underlying electronic structure. To analyse the surface electronic structure atom by atom,
we projected the density of states inside spheres around each surface atom.
The atomic-projected DOS plots shown in Fig.~\ref{fig:partial_dos} qualitatively explain the electronic effect. Ni contains an almost filled d-band. Tunneling into Ni d-states is possible only for a small unoccupied fraction above the Fermi level. All three elements Ta, W and Re exhibit a very significant local density of unoccupied d-orbitals above the Fermi level. Going from Ta$\rightarrow$W$\rightarrow$Re (increased d filling), the empty states decrease in number and shift to lower energies. Interestingly, the former effect seems to overcompensate the expected damping of tunneling into the higher-lying states, as explained in Sec.~\ref{sec:theory_contrast}. Therefore brightness contrast is dominated by the available unoccupied DOS as
speculated in our previous work for Ta in Ni(012)\cite{morgado2021revealing}.
Tunneling into these orbitals will enhance the local ionization probability and hence the FIM brightness.

\section{Conclusion}
In this work, we have laid out the foundations for an
accurate computation of tunneling-related contrast
in field ion microscopy based on state-of-the-art
density functional theory. For this, we apply the
Tersoff-Hamann approximation known from scanning tunneling
microscopy to the characteristic situation of tunneling
into imaging gas atoms hovering above the surface in
the presence of a very strong field. We
identified the numerical accuracy of wave functions from
the global solvers employed in plane wave DFT codes as a major
limitation, and developed a novel algorithm, termed EXTRA, to
recompute these tails in a very robust manner over many
orders of magnitude. Equipped with this algorithm, we
demonstrate for a prototypical case, Ta in Ni, that we
can simulate FIM contrast maps at realistic ionization
energies with practically no noise.
EXTRA allows to compute accurate DOS in the region where it is needed.
The chemical contrast between Ta and Ni is in good agreement with experiment. 
Our calculations for Ta, W, and Re suggest
that for suitable elemental combinations it should be possible to distinguish different impurities
from their relative intensities.
This new scheme paves
the way to systematically address open questions
of contrast generation in FIM. We note that the 
applicability of the EXTRA algorithm is not limited to these cases but may be
employed for other surface science questions where the tails
of the wave functions are of interest, e.g., in
overcoming tail shape limitations when localized orbitals
are used as a basis set.

\section*{Acknowledgement}
The authors like to thank Felipe M. Morgado and Baptiste Gault for their valuable discussions.
We further thank the pyiron developers, notably Marvin Poul, for their continuous technical support.
SB gratefully acknowledges financial support from the International Max Planck Research School for Sustainable Metallurgy (IMPRS SusMet).

\appendix

\section{Electrostatic potential computed via Poisson equation}
The second derivative of the electrostatic potential relates to the local charge density in Poisson's equation 
\begin{equation}
\nabla^2 \phi(\mathbf r) = -\frac{\rho(\mathbf r)}{\epsilon_0}
\label{eq:Poisson}
\end{equation}
If in-plane Fourier transform is applied along with periodic boundary conditions denoted by $\mathbf (k_x, k_y)$ where $k = \sqrt{k_x^2+k_y^2}$ and perpendicular to $z$, Eq.~(\ref{eq:Poisson}) becomes separable as 
\begin{equation}
    \frac{\partial^2}{\partial z^2} V(k_x,k_y,z) = |\mathbf k|^2 V(k_x,k_y,z) - \frac{\rho(k_x,ky,z)}{\epsilon_0}
    \;
    \label{eq:1D_seperable}
\end{equation}
The solution in the charge-free region is

(i) if $|\mathbf k| = 0$ 

\begin{equation}
    V(kx=0,ky=0,z) = D + E_z z
\end{equation}

 for constant electric field $E_z$, and

(ii) if $|\mathbf |k| > 0$ 
\begin{equation}
    V(k_x,k_y,z) = D + C_1 e^{-|\mathbf k| z}+ C_2 e^{|\mathbf k| z}
    \;.
\end{equation}

Of course, any non-zero potential variation must be matched by
charges on the counter electrode. For an ideal metallic plate-like counter electrode far away, only a homogeneous surface charge is possible that
accommodates the average field $E_z$.
Therefore, only the decaying solution $ e^{-|\mathbf k| z}$ must be considered for $|\mathbf k|>0$.

\section{Derivation of the Generalized Numerov algorithm}
\label{app:genNumerov}

The Kohn-Sham equation in three dimensions can be expressed as
\begin{equation}
    \frac{1}{2} {\partial^2 \over \partial z^2} \psi(\mathbf x,z) 
    + \left\{\epsilon - \hat V(z) \right\}\psi(\mathbf x, z) = 0
    \label{eq:Schroedinger_vhat}
\end{equation}
with the in-plane operator
\begin{equation}
    \hat V(z) = -\frac{1}{2} \left( {\partial^2 \over \partial x^2}+ {\partial^2 \over \partial y^2} \right)
    + V(\mathbf x,z)
\;.
\end{equation}                                             
$\mathbf x$ comprises the in-plane coordinates in
either real or mixed space. The in-plane operator is strictly local along the $z$ direction, but within the plane, it is semilocal (in real space) or even non-local (in mixed space).
Next, we discretize along the $z$
direction with a spacing $\zstep$, i.e., $z_n=n\zstep$, and Taylor series expansion of
$\psi$ to fifth order (with $\psi_n=\psi(\mathbf x, z_n))$:
\begin{eqnarray}
    \psi_{n\pm1} &=& \psi_n \pm \zstep\frac{\partial}{\partial z}\psi_n
                 + \frac{\zstepn^2}{2}\frac{\partial^2}{\partial z^2}\psi_n
    \nonumber\\&&
    \pm\frac{\zstepn^3}{6} \frac{\partial^3}{\partial z^3}\psi_n
                 + \frac{\zstepn^4}{24}\frac{\partial^4}{\partial z^4}\psi_n
    \nonumber\\&&
                 \pm\frac{\zstepn^5}{120} \frac{\partial^5}{\partial z^5}\psi_n
                 + \mathcal O(\zstepn^6)
                 \end{eqnarray}
Adding the equations for $\psi_{n+1}$ and $\psi_{n-1}$ makes the odd derivatives vanish.
To evaluate the term involving the 4th derivative we act on Eq.~\ref{eq:Schroedinger_vhat} with $ 1+ \frac{\zstepn^2}{12}\frac{\partial^2}{\partial z^2} $ which gives

\begin{widetext}
\begin{eqnarray}
\frac{\partial^2}{\partial z^2}\psi_n
+\frac{\zstepn^2}{12} \frac{\partial^4}{\partial z^4}\psi_n +\left\{\epsilon - \hat V(z) \right\}\psi_n+\frac{\zstepn^2}{12} \frac{\partial^2}{\partial z^2}\left\{\epsilon - \hat V(z) \right\}\psi_n=0
\end{eqnarray}
Substituting for $ \frac{\partial^2}{\partial z^2}\psi_n
+\frac{\zstepn^2}{12} \frac{\partial^4}{\partial z^4}\psi_n$ and evaluate $\frac{\partial^2}{\partial z^2}\left\{\epsilon - \hat V(z) \right\}\psi_n$ gives

\begin{eqnarray}
\frac{\partial^2}{\partial z^2}\left\{\epsilon - \hat V(z_n) \right\}\psi_n = \frac{\left\{\epsilon - \hat V(z_{n+1}) \right\}\psi_{n+1}+\left\{\epsilon - \hat V(z_{n-1}) \right\}\psi_{n-1}-2\left\{\epsilon - \hat V(z_n) \right\}\psi_n}{\zstepn^2}
\end{eqnarray}

With this substitution, one can rearrange terms to arrive at

\begin{equation}
\left[1+\frac{1}{6}\zstepn^2\left\{\epsilon-\hat{V}(z_{n+1})\right\}\right]\psi_{n+1} = 2\left[1-\frac{5}{6}\zstepn^2\left\{\epsilon-\hat{V}(z_{n})\right\}\right]\psi_n
-\left[1+\frac{1}{6}\zstepn^2\left\{\epsilon-\hat{V}(z_{n-1})\right\}\right]\psi_{n-1}
\label{eq:gen_Numerov_final}
\end{equation}
\end{widetext}

%

\end{document}